\documentclass[times,astrobib,amssymb,usenatbib]{mn2e} 
\usepackage{times} 
\usepackage{latexsym,amsmath,amssymb,amsfonts} 
\usepackage{graphicx} 
\usepackage{fancyhdr}  
\usepackage{natbib}  
\usepackage{supertabular}
\usepackage{longtable}
\usepackage{url}
\usepackage{dcolumn}
\usepackage{morefloats}	 
\usepackage{rotating}  	 
\usepackage{tikz}    	 
\usepackage{adjustbox}
\usepackage{blindtext}  
\usepackage[breaklinks,colorlinks,citecolor=blue,linkcolor=magenta]{hyperref}   
\begin{document}
\newcommand{\sqcm}{cm$^{-2}$}  
\newcommand{\lya}{Ly$\alpha$}
\newcommand{\lyb}{Ly$\beta$}
\newcommand{\lyg}{Ly$\gamma$}
\newcommand{\lyd}{Ly$\delta$}
\newcommand{\HeI}{\mbox{He\,{\sc i}}}
\newcommand{\HeII}{\mbox{He\,{\sc ii}}}
\newcommand{\HI}{\mbox{H\,{\sc i}}}
\newcommand{\HII}{\mbox{H\,{\sc ii}}}
\newcommand{\HH}{\mbox{H{$_2$}}}
\newcommand{\hh}{\mbox{\tiny H{$_2$}}} 
\newcommand{\hi}{\mbox{\tiny H{\sc i}}} 
\newcommand{\OI}{\mbox{O\,{\sc i}}}
\newcommand{\OII}{\mbox{O\,{\sc ii}}}
\newcommand{\OIII}{\mbox{O\,{\sc iii}}}
\newcommand{\OIV}{\mbox{O\,{\sc iv}}}
\newcommand{\OV}{\mbox{O\,{\sc v}}}
\newcommand{\OVI}{\mbox{O\,{\sc vi}}}
\newcommand{\OVII}{\mbox{O\,{\sc vii}}}
\newcommand{\OVIII}{\mbox{O\,{\sc viii}}}
\newcommand{\CaVIII}{\mbox{Ca\,{\sc viii}}}
\newcommand{\CaVII}{\mbox{Ca\,{\sc vii}}}
\newcommand{\CaVI}{\mbox{Ca\,{\sc vi}}}
\newcommand{\CaV}{\mbox{Ca\,{\sc v}}}
\newcommand{\CIV}{\mbox{C\,{\sc iv}}}
\newcommand{\CV}{\mbox{C\,{\sc v}}}
\newcommand{\CVI}{\mbox{C\,{\sc vi}}}
\newcommand{\CII}{\mbox{C\,{\sc ii}}}
\newcommand{\CI}{\mbox{C\,{\sc i}}}
\newcommand{\CIIs}{\mbox{C\,{\sc ii}}$^\ast$}
\newcommand{\CIII}{\mbox{C\,{\sc iii}}}
\newcommand{\SiI}{\mbox{Si\,{\sc i}}}
\newcommand{\SiII}{\mbox{Si\,{\sc ii}}}
\newcommand{\SiIII}{\mbox{Si\,{\sc iii}}}
\newcommand{\SiIV}{\mbox{Si\,{\sc iv}}}
\newcommand{\SiXII}{\mbox{Si\,{\sc xii}}}
\newcommand{\SV}{\mbox{S\,{\sc v}}}
\newcommand{\SIV}{\mbox{S\,{\sc iv}}}
\newcommand{\SIII}{\mbox{S\,{\sc iii}}}
\newcommand{\SII}{\mbox{S\,{\sc ii}}}
\newcommand{\SI}{\mbox{S\,{\sc i}}}
\newcommand{\ClI}{\mbox{Cl\,{\sc i}}}
\newcommand{\ArI}{\mbox{Ar\,{\sc i}}}
\newcommand{\NI}{\mbox{N\,{\sc i}}}
\newcommand{\NII}{\mbox{N\,{\sc ii}}}
\newcommand{\NIII}{\mbox{N\,{\sc iii}}}
\newcommand{\NIV}{\mbox{N\,{\sc iv}}}
\newcommand{\NV}{\mbox{N\,{\sc v}}}
\newcommand{\PV}{\mbox{P\,{\sc v}}}
\newcommand{\PII}{\mbox{P\,{\sc ii}}}
\newcommand{\PIII}{\mbox{P\,{\sc iii}}}
\newcommand{\PIV}{\mbox{P\,{\sc iv}}}
\newcommand{\NeVIII}{\mbox{Ne\,{\sc viii}}}
\newcommand{\ArVIII}{\mbox{Ar\,{\sc viii}}}
\newcommand{\NeV}{\mbox{Ne\,{\sc v}}}
\newcommand{\NeVI}{\mbox{Ne\,{\sc vi}}}
\newcommand{\NeX}{\mbox{Ne\,{\sc x}}} 
\newcommand{\NaIX}{\mbox{Na\,{\sc ix}}} 
\newcommand{\MgII}{\mbox{Mg\,{\sc ii}}}
\newcommand{\FeII}{\mbox{Fe\,{\sc ii}}}
\newcommand{\MgX}{\mbox{Mg\,{\sc x}}}
\newcommand{\AlXI}{\mbox{Al\,{\sc xi}}}
\newcommand{\FeIII}{\mbox{Fe\,{\sc iii}}}
\newcommand{\NaI}{\mbox{Na\,{\sc i}}}
\newcommand{\CaII}{\mbox{Ca\,{\sc ii}}}
\newcommand{\zabs}{$z_{\rm abs}$}
\newcommand{\zmin}{$z_{\rm min}$}
\newcommand{\zmax}{$z_{\rm max}$}
\newcommand{\zqso}{$z_{\rm qso}$}
\newcommand{\subHe}{_{\it HeII}}
\newcommand{\subH}{_{\it HI}}
\newcommand{\subHLy}{_{\it H Ly}}
\newcommand{\degree}{\ensuremath{^\circ}}
\newcommand{\lapp}{\mbox{\raisebox{-0.3em}{$\stackrel{\textstyle <}{\sim}$}}}
\newcommand{\gapp}{\mbox{\raisebox{-0.3em}{$\stackrel{\textstyle >}{\sim}$}}}
\newcommand{\be}{\begin{equation}}
\newcommand{\en}{\end{equation}}
\newcommand{\di}{\displaystyle}
\def\tworule{\noalign{\medskip\hrule\smallskip\hrule\medskip}} 
\def\onerule{\noalign{\medskip\hrule\medskip}} 
\def\bl{\par\vskip 12pt\noindent}
\def\bll{\par\vskip 24pt\noindent}
\def\blll{\par\vskip 36pt\noindent}
\def\rot{\mathop{\rm rot}\nolimits}
\def\alf{$\alpha$}
\def\refff{\leftskip20pt\parindent-20pt\parskip4pt}
\def\zabs{$z_{\rm abs}$}
\def\zqso{$z_{\rm qso}$}
\def\zem{$z_{\rm em}$}
\def\mgii{Mg\,{\sc ii}~} 
\def\feiia{Fe\,{\sc ii}$\lambda$2600}
\def\mgia{Mg\,{\sc i}$\lambda$2852}
\def\mgiia{Mg\,{\sc ii}$\lambda$2796}
\def\mgiib{Mg\,{\sc ii}$\lambda$2803}
\def\mgiiab{Mg\,{\sc ii}$\lambda\lambda$2796,2803}
\def\wobs{$w_{\rm obs}$}
\def\kms{km~s$^{-1}$}
\def\bnt{$b_{\rm nt}$}
\def\fosc{$f_{\rm osc}$}
\def\chisq{$\chi^{2}$}
\def\dtype{$\delta_{\rm type}$}  
\title[An EUV mini-BAL variability]{ 
\vskip-0.5cm  
On the covering fraction variability in an EUV mini-BAL outflow from PG~1206+459} 
\author[S. Muzahid et al.]
{
\parbox{\textwidth}{\vskip-0.5cm  
S. Muzahid$^{1}$\thanks{E-mail: sowgatm@gmail.com},     
R. Srianand$^{2}$,  
J. Charlton$^{1}$, and 
M. Eracleous$^{1,3}$      
} 
\vspace*{4pt}\\   
$^{1}$ Department of Astronomy \& Astrophysics, The Pennsylvania State University, 525 Davey Lab, University Park, State College, PA 16802, USA \\ 
$^{2}$ Inter-University Centre for Astronomy and Astrophysics, Post Bag 4, Ganeshkhind, Pune 411007, India \\ 
$^{3}$ Institute for Gravitation and the Cosmos, The Pennsylvania State University, University Park, State College, PA 16802, USA \\ 
}   
\date{\vskip-1cm Accepted. Received; in original form}
\maketitle
\label{firstpage}
\begin {abstract} 
We report on the first detection of extreme-ultraviolet (EUV) absorption variability in the \NeVIII\ $\lambda\lambda$770,780 mini-broad absorption line (mini-BAL) in the spectrum of the quasar (QSO) PG~1206+459. The observed equivalent width (EW) of the \NeVIII\ doublet shows a $\sim$4$\sigma$ variation over a timescale of 2.8 months in the QSO's rest-frame. Both members of the \NeVIII\ doublet exhibit non-black saturation, indicating partial coverage of the continuum source. An increase in the \NeVIII\ covering fraction from $f_c = 0.59\pm0.05$ to $0.72\pm0.03$ is observed over the same period. The \NeVIII\ profiles are too highly saturated to be susceptible to changes in the ionization state of the absorbing gas. In fact, we do not observe any significant variation in the EW and$/$or column density after correcting the spectra for partial coverage. We, thus, propose transverse motions of the absorbing gas as the cause of the observed variability. Using a simple model of a transiting cloud we estimate a transverse speed of $\sim$1800~\kms. For Keplerian motion, this corresponds to a distance between the absorber and the central engine of $\sim$1.3~pc, which places the absorber just outside the broad-line region. We further estimate a density of $\sim$5$\times10^{6}$~cm$^{-3}$ and a kinetic luminosity of $\sim$$10^{43}$--$10^{44}$~erg~s$^{-1}$. Such large kinetic powers suggest that outflows detected via EUV lines are potentially major contributors to AGN feedback.   
\end {abstract}
\vskip-1cm 
\section{Introduction} 
\label{sec:intro}  

Broad absorption line (BAL; line spread, $\Delta v \sim$few$\times$1000~\kms) and mini-BAL ($\Delta v \sim$few$\times$100~\kms) systems detected via UV absorption troughs blue-shifted with respect to the QSO's emission redshift are thought to originate in an equatorial wind that is launched from the accretion disk near the central supermassive black hole \citep[SMBH; e.g.,][]{Murray95a,Arav95,Proga00}. The mass, momentum, and energy carried by such winds are increasingly invoked as the primary feedback mechanisms to explain the evolution of SMBH \citep[]{Silk98,King03,King05} and their host-galaxies \citep[]{Begelman05,DiMatteo05}. For example, it has been shown that an outflow with a kinetic power of a few per cent of the Eddington luminosity provides sufficient feedback to quench star formation, thus coupling the evolution of the host-galaxy and the growth of the central SMBH \citep[]{DiMatteo05}. QSO outflows can also be responsible for metal enrichment in the interstellar medium (ISM), in the circumgalactic medium (CGM), and even in the intergalactic medium \citep[IGM;][]{Wiersma11}. Therefore, the study of QSO outflows via BAL and mini-BAL systems is immensely important. In particular, the incidence rate, outflow velocity, variability timescale, and covering fraction ($f_c$) of the absorption troughs provide essential clues about the location, geometry, and energetics of the flow.  

It is now well known that BAL systems show variability in absorption profiles over a time period of 1--10 yrs in the QSO rest-frame \citep[e.g.,][and references therein]{Barlow94,Srianand01,Lundgren07,Gibson08,Capellupo11,FilizAk13,Vivek12,Vivek14}. \cite{FilizAk13} found a BAL fraction of $\sim$50--60 per cent that exhibit variability over time-scales of 1--4 yrs in the QSO's rest-frame. Additionally, BAL variability is found to be larger for longer timescales \citep[e.g.,][]{Gibson08,Capellupo11,Vivek14}. For example, \cite{Capellupo11} have found variability in $\sim$40 per cent of BAL absorbers on timescales of $<1$ yr with that fraction increasing to $\sim$65 per cent on timescales of 4--8 yr in the QSO rest-frame. BAL variability is generally explained by: (a) change in $f_c$ due to motions of the absorber across the line of sight and$/$or (b) change in the ionization state of the gas. In either case, it is possible to determine the dynamics and location of the absorber with respect to the central engine under some assumptions. Note that the above explanations hold for mini-BAL variability as well.   

While BAL variability is well studied via several ongoing monitoring campaigns by different groups with a fairly large number of objects, very little is known for mini-BALs. In fact, variability in mini-BAL absorption is studied only in a handful of cases \citep[i.e.,][]{Hamann97vary,Narayanan04,Misawa07a,Hidalgo11,Misawa14}. Interestingly, \cite{Misawa07a} have demonstrated that the aforementioned causes for line variability cannot reproduce the rapid, simultaneous variation in multiple troughs of the mini-BAL system detected towards HS~1603+3820. Instead, they proposed that a clumpy ``screen" of gas with variable optical depth and covering fraction, located between the continuum source and the absorbing gas, mimics the variation in the ionizing continuum. The variability study of a sample of seven mini-BALs was recently presented by \cite{Misawa14}. Assuming that the observed variations result from changes in the ionization state of the mini-BAL gas, they have inferred lower limits to the gas density of $\sim$$10^{3}-10^{5}$ cm$^{-3}$, corresponding to upper limits on the distance of the absorbers from the central engine of $\sim$~few kpc. Note that unlike BAL systems, absorption line parameters can be more robustly determined for mini-BAL systems because (a) the QSO continuum is easily identifiable even in the presence of mini-BALs, (b) doublets are not self-blended due to relatively narrow line profiles, and (c) the kinematic structure is often discernible due to lack of heavy line-saturation \citep[see e.g.,][]{Muzahid12b}. Mini-BAL systems are thus excellent tools for measuring outflow properties, e.g., density, metallicity, outflow mass, mass-flow rate, and kinetic luminosity that are essential for quantifying the so called ``AGN feedback" \citep[]{Arav13}. 

Recently, in an archival {\it Hubble Space Telescope} ($\it HST$)$/$ Cosmic Origins Spectrograph (COS) study, we have reported a new population of associated absorbers detected via absorption from EUV ions (e.g., \NeVIII, \NaIX, and \MgX) in the spectra of intermediate-$z$ QSOs \citep[][hereafter M13]{Muzahid13}. We have found that a significant fraction ($\sim$40 per cent) of the intermediate-$z$ QSOs exhibit outflows observable via EUV lines. These absorbers are very highly ionized with ionization parameters (i.e. 0.5$\lesssim \log U \lesssim$1.0) consistent with their being intermediate to X-ray ``warm absorbers" \citep[]{Halpern84} and BALs. The velocity spreads (100--800 \kms) of the \NeVIII\ absorption lines suggest that they are mini-BALs. 
Here we present a fortuitous case from the sample presented in M13 in which the QSO (PG~1206+459) was observed at three different epochs. PG~1206+459 is an UV-bright QSO with a $V$-band magnitude of $\sim$15.4 and has an emission redshift of \zem~$=$~1.163. The multi-epoch spectroscopic observations of the QSO allowed us to study variability of the \OIV, \NIV, and \NeVIII\ absorption. We found a $\sim$4$\sigma$ variation in the \NeVIII\ equivalent width. This is the first case of EUV mini-BAL variability reported so far. This article is organized as follows: In Section~\ref{sec:obs} we present observations and data reduction of the QSO PG~1206+459. Absorption line analysis and the QSO's properties are presented in Section~\ref{sec:ana}. In Section~\ref{sec:diss} we discuss the implications of our findings. In Section~\ref{sec:summ} we briefly summarized our main results. Throughout this article we adopt a flat $\Lambda$CDM cosmology with $\Omega_{\Lambda} = 0.73$, $\Omega_{\rm M} = 0.27$, and $H_{0} = 71$~\kms~Mpc$^{-1}$. Unless specified otherwise, the EWs are given in the observed frame whereas the timescales are in the QSO's rest-frame.

\begin{table}
\caption{Details of the $HST/$COS observations.}       
\begin{tabular}{m{0.7cm}m{0.5cm}m{0.3cm}m{0.7cm}m{1.4cm}m{0.9cm}m{0.5cm}m{0.5cm}} 
\hline  
Grating$^{a}$ & $\lambda_{\rm cen}^{b}$  & Tilt$^{c}$ & $t_{\rm exp}^{d}$  & Date of      &      MJD$^{e}$  & Epoch &  $S/N^{f}$  \\ 
              &      (\AA)               &            &   (sec)            & Observation  &                 &       &             \\ \hline         
G160M  &  1600  &    1  &    2361.2 & 2009-12-29     & 55194.5 &  1   &    \\ 
G160M  &  1600  &    2  &    3137.2 & 2009-12-29     & 55194.5 &  1   &    \\ 
G160M  &  1623  &    3  &    3137.2 & 2009-12-29     & 55194.5 &  1   &    \\ 
G160M  &  1623  &    4  &    3137.2 & 2009-12-29     & 55194.5 &  1   &  11 \\                                   
G160M  &  1600  &    1  &    3137.2 & 2010-01-05     & 55201.5 &  2   &    \\ 
G160M  &  1600  &    2  &    3137.2 & 2010-01-05     & 55201.5 &  2   &    \\ 
G160M  &  1600  &    3  &    3137.2 & 2010-01-05     & 55201.5 &  2   &    \\ 
G160M$^{1}$  &  1600  &    4  &    3137.0 & 2010-01-05     & 55201.5 &  2   & 10 \\   
                                                    
G160M  &  1623  &    1  &    2361.2 & 2010-06-14     & 55361.5 &  3   &    \\ 
G160M  &  1623  &    2  &    3137.1 & 2010-06-14     & 55361.5 &  3   &    \\ 
G160M  &  1623  &    3  &    3137.2 & 2010-06-14     & 55361.5 &  3   &    \\ 
G160M  &  1623  &    4  &    3137.2 & 2010-06-14     & 55361.5 &  3   &  12 \\    
G130M  &  1309  &    1  &    2406.2 & 2010-01-04     & 55200.5 &  2  &     \\ 
G130M  &  1309  &    3  &    3137.2 & 2010-01-04     & 55200.5 &  2  &     \\ 
G130M  &  1309  &    4  &    3137.2 & 2010-01-04     & 55200.5 &  2  &     \\ 
G130M  &  1327  &    1  &    3137.2 & 2010-01-04     & 55200.5 &  2  &     \\ 
G130M  &  1327  &    3  &    3137.2 & 2010-01-04     & 55200.5 &  2  &     \\ 
G130M  &  1327  &    4  &    2406.2 & 2010-01-05     & 55201.5 &  2  &  27 \\ 
\hline  
\end{tabular}        
\label{tab:obs}  
~\\  
Notes-- $^{a}$Grating used. 
$^{b}$Central wavelength.  
$^{c}$Fixed-pattern offset position. 
$^{d}$Exposure time in second. 
$^{e}$Modified Julian date corresponding to the observation date.   
$^{f}$$S/N$ per resolution element of the coadded data at $\sim$1575~\AA\    
for G160M grating and at $\sim$1230~\AA\ for G130M grating.       
$^{1}$This exposure ($lb1o05kqq$) was bad and not used for co-addition.       
\end{table}  		  	
  

\section{Observations and Data Reduction}  
\label{sec:obs} 

Far-UV (FUV) spectra of the UV-bright QSO PG~1206+459 (\zem\ = 1.163) were obtained using $HST/$COS Cycle-17 observations, under program ID: GO-11741 (PI: Todd Tripp). These observations consist of G130M and G160M grating integrations at a medium spectral resolution of $R\sim$20,000 (FWHM~$\sim$18~\kms). The data were retrieved from the $HST$ archive and reduced using the STScI CALCOS v2.21 pipeline software. The reduced data were flux calibrated. To increase the spectral signal-to-noise ratio ($S/N$), spectra from individual exposures were aligned and coadded using the IDL code ``coadd\_x1d" developed by \cite{Danforth10}. The properties of COS and its in-flight operations are discussed by \cite{Osterman11} and \cite{Green12}. The coadded spectra are significantly oversampled, with six raw pixels per resolution element. We thus binned the data by six pixels. This improves the spectral $S/N$ per pixel by a factor of $\sqrt6$. All our measurements and analyses were performed on the binned data. The binning does not affect our EW and$/$or $f_c$ measurements since the absorption lines of interest are much broader than the spectral resolution. Continuum normalization was done by  fitting the  line-free regions with  smooth low-order polynomials. Since there is no \lya\ forest crowding at low-$z$ and the absorption lines of interest do not fall on top of any emission lines, the continua were very well defined in most part. COS wavelength calibration is known to have uncertainties at the level of 10--15~\kms\ \citep[]{Savage11,Meiring13}. Using molecular hydrogen absorption lines, \cite{Muzahid15a} have noted that while in majority of the cases wavelengths are accurate within $\sim \pm$5~\kms, large offsets ($>$10~\kms) are often noticeable at the edges of the COS spectra. We noticed that the wavelength calibration uncertainty is usually $<$5~\kms\ for PG~1206+459. However, the uncertainty increases to $\sim$10~\kms\ at the edge of the spectra as also noted by \cite{Muzahid15a}.    

Besides COS, we have used the $HST/$Space Telescope Imaging Spectrograph (STIS) spectrum of PG~1206+459. The STIS spectrum was obtained using the E230M grating under program ID: 8672 (PI: Chris Churchill). We refer the reader to \cite{Ding03} for the full details on the STIS observations. The spectrum is only used for calculating rest-frame 1350~\AA\ flux.

\begin{figure*} 
\centerline{\vbox{ 
\centerline{\hbox{ 
\includegraphics[width=0.90\textwidth,angle=00]{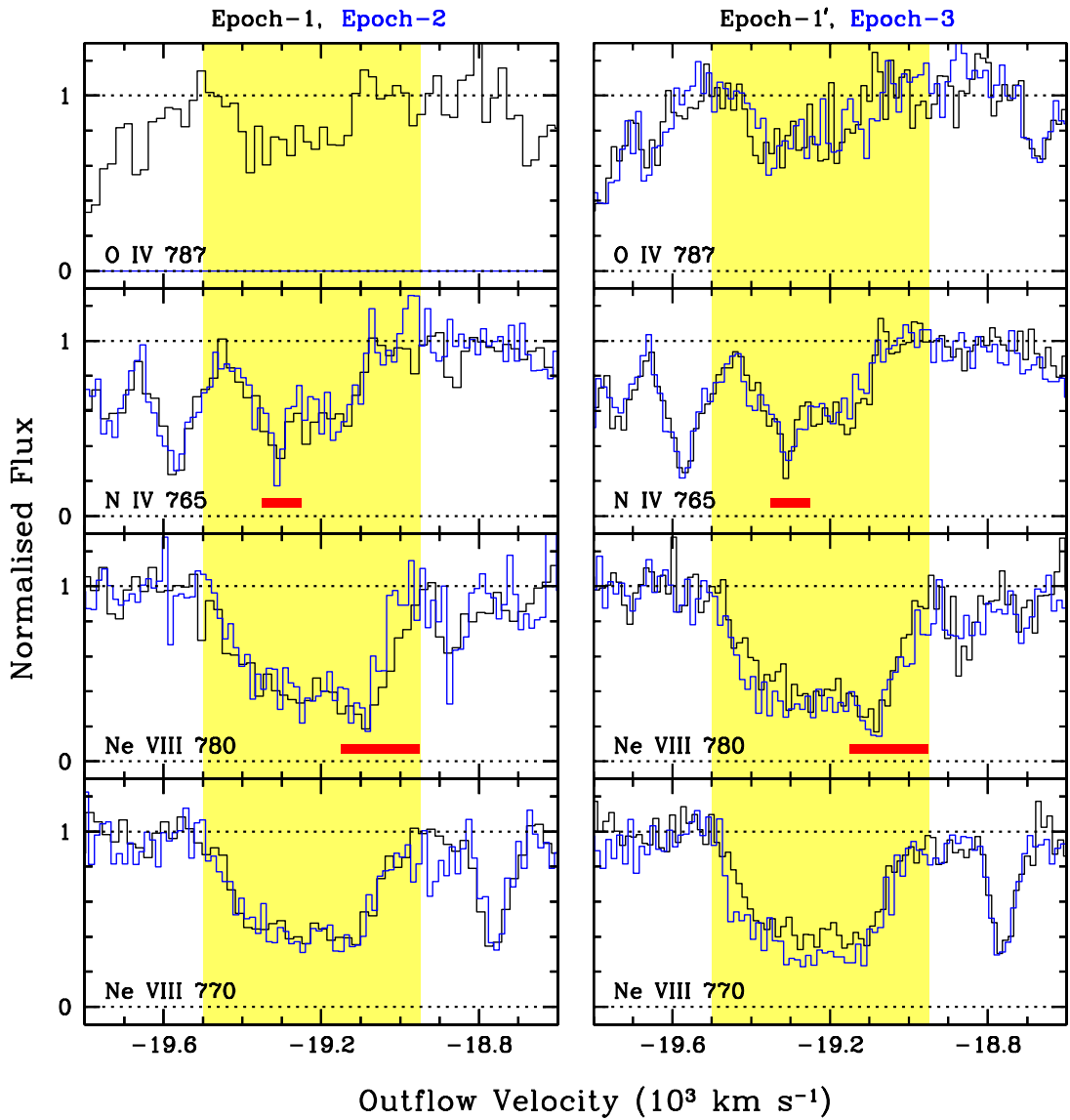}   
}}
}}
\vskip-0.35cm  
\caption{Normalised flux against the outflow velocity, measured with respect to the \zem~$=$~1.163, for different ions. The absorption from the outflow are shown by the shaded regions. All other absorption are unrelated and intervening. The blended pixels are marked by the horizontal red bars. In each panel the data with different colors represent different epochs as indicated in the top. No significant variation is seen in the left panel between Epoch-1 and Epoch-2 for any of the troughs. We therefore co-add the Epoch-1 and Epoch-2 data to obtain a better $S/N$ spectrum which is represented by Epoch-$1^\prime$ in the right panel. While no appreciable variation is seen between Epoch-$1^\prime$ and Epoch-3 in the \NIV\ and \OIV\ absorption, the \NeVIII\ doublet shows a significant change in the absorption troughs. Note that only one central wavelength setting was used for the G160M grating observations in Epoch-2 and the \OIV\ line falls in the gap of the Epoch-2 spectrum. 
}           
\label{fig:vary}  
\end{figure*}   

In Table~\ref{tab:obs} we summarize the details of the observations. It is clear from the table that the G160M observations were taken in three different epochs. Epoch-1 and Epoch-2 are separated by $\sim$7 days, corresponding to $\sim$3.2 days in the QSO's rest-frame. Epoch-1 and Epoch-3 are separated by $\sim$6 months ($\sim$2.8 months in the QSO's rest-frame). We note that the G160M spectrum covers only the \NIV~$\lambda$765, \OIV~$\lambda$~787, and \NeVIII~$\lambda\lambda$770,780 transitions from the mini-BAL outflow we studied here. Unfortunately, other important high-ionization species (e.g., \NaIX~$\lambda\lambda$681,694, \MgX~$\lambda\lambda$609,624) fall in the wavelength range covered by the G130M grating and are observed only once.

\section{Analysis}   
\label{sec:ana}

\subsection{Properties of the absorber}  
\label{subsec:ABSprop}    

The mini-BAL outflow from QSO PG~1206+459 (\zem~$=$~1.163) was first reported in M13 as a part of our $HST/$COS survey of intrinsic absorbers detected via \NeVIII~$\lambda\lambda$770,780 doublet transitions. The absorption redshift of \zabs~$=$~1.02854 corresponds to an outflow velocity of $\sim$19,200~\kms. Both members of the \NeVIII\ doublet are detected with $\rm EW/\sigma_{EW}>20$. \linebreak The \NeVIII\ absorption is spread over $\Delta v\sim$360~\kms\ with a total column density of $N(\NeVIII)>10^{16}$~\sqcm. A very weak \lya\ line is detected with $\log (N(\HI)/{\rm cm^{-2}})\lesssim$14 in the $HST/$STIS spectrum. Interestingly, no singly or doubly ionized species are detected from this absorber. Besides \NeVIII, we have reported the presence of several other high-(\ArVIII, \NaIX, \MgX) and low-(\OIV, \NIV, \OV, \NV) ionization species (see Fig.~7 of M13). But only \NIV~$\lambda$765, \NeVIII~$\lambda$770,780, and \OIV~$\lambda$787 lines have multi-epoch coverage.

In Fig.~\ref{fig:vary} we show the absorption profiles of \NIV~$\lambda$765, \OIV~$\lambda$~787, and \NeVIII~$\lambda\lambda$770,780 transitions for the three different epochs. No significant variation in any of these absorption troughs between Epoch-1 and Epoch-2 (separated by $\sim$3.2 days in the QSO's rest-frame) is apparent from the figure. To quantify the degree of variability in an absorption, we measure EWs at different epochs and define absorption variability significance as follows:  
\begin{equation}  
S_{\rm ij} = \rm{\frac{EW_j - EW_i}{\sqrt{2(\sigma_{EW_j}^2+ \sigma_{EW_i}^2)}}}~,   
\label{eqn:sig}   
\end{equation} 
where, $\rm EW_j$ and $\sigma_{\rm EW_j}$ are the equivalent width and its error, respectively, as measured in j-th epoch. Note that a positive $S_{\rm ij}$ indicates an increase in EW at later epoch. We have estimated that the error in the observed EW due to the continuum placement uncertainty is on the order of $\sigma_{\rm EW}$. We, thus, introduce the $\sqrt2$ factor in the denominator in Eq.~\ref{eqn:sig} to take care of the continuum placement uncertainty \citep[see also][]{Misawa14}.    

\begin{figure} 
\centerline{\vbox{ 
\centerline{\hbox{ 
\includegraphics[width=0.51\textwidth,angle=00]{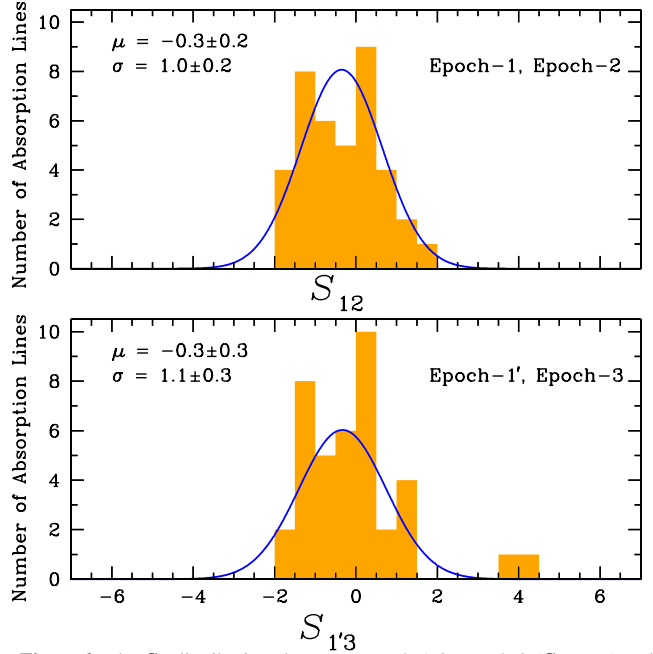}   
}}
}}
\vskip-0.35cm  
\caption{The $S_{\rm ij}$ distributions between Epoch-1 \& Epoch-2 ($S_{12}$, top) and between Epoch-1$^{\prime}$ \& Epoch-3 ($S_{1^\prime3}$, bottom) for a set of 39 strong absorption lines detected in all three epochs (see text). The details of these absorption lines are summarized in Appendix~\ref{app:EWlist}. Each of the observed $S_{\rm ij}$ distributions is modelled as Gaussian with fit parameters given in each panel. No absorption lines shows a $>$2$\sigma$ variability between any of the epochs except for the \NeVIII~$\lambda\lambda$770,780 transitions which show $\sim$4$\sigma$ variability, in the bottom panel, between Epoch-1$^{\prime}$ \& Epoch-3.    
}             
\label{fig:sigma}   
\end{figure}

Next, we choose a set of 39 strong absorption lines that are detected with $\rm EW/\sigma_{EW}>5$ in all three epochs in the wavelength ranges 1440--1587~\AA\ and 1623--1775~\AA\footnote{We do not use lines within $\sim$20~\AA\ from both blue- and red-end of the spectra due to wavelength calibration uncertainty.}. Details of these lines are given in Appendix~\ref{app:EWlist}. Note that all but three of these lines, i.e., NIV~$\lambda765$, \NeVIII~$\lambda$770 and $\lambda$780 lines from the mini-BAL\footnote{\OIV~$\lambda$787 is not considered here since it is not covered in Epoch-2 data.}, are intervening and should not show any considerable variation with time. For all these absorption lines we have calculated $S_{\rm ij}$. The top panel of Fig.~\ref{fig:sigma} shows the distribution of $S_{\rm ij}$ as measured between Epoch-1 and Epoch-2 ($S_{12}$). It is apparent from the figure that none of the absorption lines, including the three from the mini-BAL system, shows variability with a significance of $>$2$\sigma$. A Gaussian fit to the {$S_{12}$} distribution gives a mean of $\mu=-0.3\pm0.2$ and a standard deviation of $\sigma=1.0\pm0.2$. A 2$\sigma$ variation, thus, is not significant for the data we have. Since there is no considerable variation in any absorption we co-add the Epoch-1 and Epoch-2 data in order to increase the spectral S$/$N. The resultant data, with marginally improved S$/$N of $\sim$14 per resolution element, are shown as Epoch-$1^\prime$ in Fig.~\ref{fig:vary}. The Epoch-$1^{\prime}$ spectrum has spectral coverage of 1150--1800~\AA. For interested readers we have shown the Epoch-$1^{\prime}$ spectrum in Appendix~\ref{app:fullspec}.

In the bottom panel of Fig.~\ref{fig:sigma} we show the $S_{\rm ij}$ distribution as calculated between Epoch-$1^\prime$ and Epoch-3 ($S_{1^\prime3}$). Here also we find that none, but two of the absorption lines, shows variability with a significance of $>$2$\sigma$. A Gaussian fit to the $S_{1^\prime3}$ distribution results in similar $\mu$ and $\sigma$ values as we obtained for the $S_{12}$ distribution. The two absorption lines showing $\sim$4$\sigma$ variation are the \NeVIII~$\lambda$770 and $\lambda$780 of the mini-BAL outflow (see Appendix~\ref{app:EWlist}). Clearly, the variations are statistically significant and cannot be attributed to some systematic uncertainties. The \NIV\ line, however, does not show any considerable variability between Epoch-$1^\prime$ and Epoch-3, e.g., shows only a $-0.4\sigma$ variation. Moreover, the \OIV\ line shows only a $\sim$0.2$\sigma$ variation between Epoch-1 and Epoch-3.          

\begin{figure} 
\centerline{\vbox{ 
\centerline{\hbox{ 
\includegraphics[width=0.5\textwidth,angle=00]{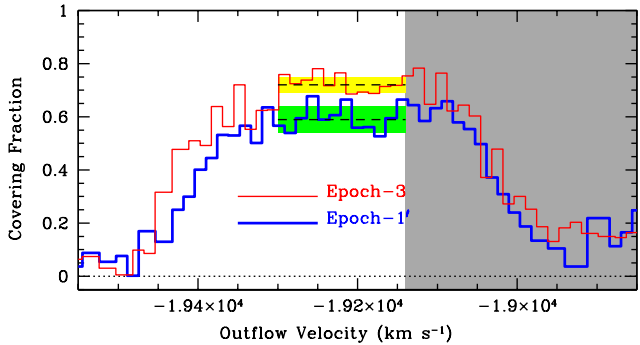}  
}}
}}
\vskip-0.35cm 
\caption{The \NeVIII\ covering fraction against the outflow velocity (relative to the \zem) 
 for the two different epochs. The thick blue curve represents Epoch-$1^\prime$, whereas 
the thin red curve is for Epoch-3. The median $f_c$ values as shown by the horizontal  
dashed lines exhibit an increase from $0.59\pm0.05$ (std) to $0.72\pm0.03$ (std) from the 
Epoch-$1^\prime$ to Epoch-3. The height of the shaded horizontal bars represent the $1\sigma$ 
range in the measured $f_c$ values. Note that the $f_c$ measurements in the grey shaded region 
are not reliable due to an unknown blend in the red wing of the \NeVIII~$\lambda$780 absorption.  
}   
\label{fig:covf}   
\end{figure}

In a situation when an absorber covers the background continuum emitting region only partially with a covering fraction of $f_c$, then the observed and ``true" optical depths ($\tau_{\lambda}$) are related via:   
\begin{equation}  
\exp{(-\tau^{\rm obs}_{\lambda})}\equiv R_{\lambda}^{\rm obs} =(1-f_c)+f_c~\exp(-\tau^{\rm true}_{\lambda})~.  
\label{eqn:flux} 
\end{equation}  
Here $R_{\lambda}^{\rm obs}$ is the observed normalized flux. Here we assume that the background source is partially covered by a homogeneous absorber with a constant $f_c$. However, $f_c$ can be strongly dependent on velocity along the absorption trough \citep[e.g.,][]{Arav08}. In the case of a doublet with rest-frame wavelengths of $\lambda_1$ and $\lambda_2$, and oscillator strengths of $f_1$ and $f_2$, the normalized fluxes of $R_1^{\rm obs}$ and $R_2^{\rm obs}$ at any velocity with respect to the line centroid are related by:  
\begin{equation}  
R_2^{\rm obs}(v) = (1-f_c)+f_c\times\left(\frac{R_1^{\rm obs}(v)-1+f_c}{f_c}\right)^{\gamma}~,  
\label{eqn:fc}  
\end{equation}     
where, $\gamma = f_2\lambda_2/f_1\lambda_1$, which is close to 2 for the \NeVIII\ doublet \citep[see e.g.,][]{Petitjean99,Srianand99}. Using Eq.~\ref{eqn:fc} we have calculated the \NeVIII\ covering fraction as a function of outflow velocity as shown in Fig.~\ref{fig:covf} \citep[see also][]{Muzahid12b,Muzahid13}. The median $f_c$ value obtained for the Epoch-$1^\prime$ data is $0.59\pm0.05$, whereas it is $0.72\pm0.03$ for the Epoch-3 data. The error bars correspond to the standard deviations of $f_c$ values measured in each pixel. Note that the median $f_c$ values are determined from the core of the absorption, i.e., only using the pixels with highest optical depths marked by the horizontal bars. Nevertheless, it is evident from the figure that the Epoch-3 data show higher $f_c$ values as compared to the Epoch-$1^\prime$ data in almost every pixel over the entire velocity range.   

Finally, in order to evaluate whether the \NeVIII\ column density has increased with time, we perform EW measurements in the covering fraction corrected spectra (i.e. the spectra obtained by inverting the Eq.~\ref{eqn:flux}). For both the Epoch-$1^\prime$ and Epoch-3 data we measure EW$_{\rm obs}$(\NeVIII~$\lambda$770)~=~$2.00\pm0.04$ \AA. Thus, no significant change in \NeVIII\ equivalent width is noticeable in the partial coverage corrected spectra. A change in $N(\NeVIII)$ without a significant change in EW is possible in the flat part of the curve-of-growth. However the fact that we do not see any change in the wings of the profile indicates no considerable change in $N(\NeVIII)$ over the course of 2.8 months in the QSO's rest-frame. 

\subsection{Properties of the background QSO}    
\label{subsec:QSOprop}     

\begin{figure} 
\centerline{\vbox{ 
\centerline{\hbox{ 
\includegraphics[width=0.5\textwidth,angle=00]{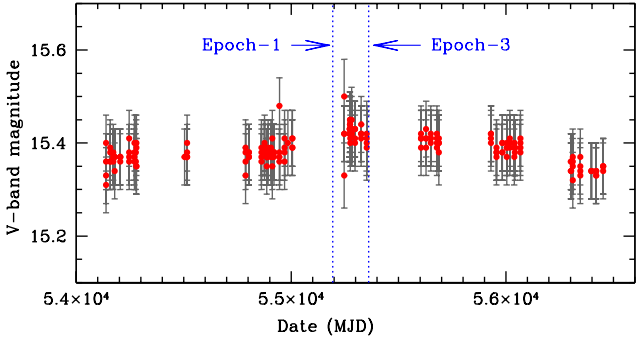}  
}}
}}
\vskip-0.35cm 
\caption{The CRTS $V$-band magnitudes of QSO PG~1206+459 plotted against the modified 
Julian date (MJD). Clearly, the background source shows a negligible ($<0.1$ mag) variation 
over the entire MJD range. The Epoch-1 and Epoch-3 are indicated by the arrows.}   
\label{fig:lightcurve}   
\end{figure}

The UV bright QSO PG~1206+459 has a $V$-band magnitude of $\sim$15.4. The Catalina Real-Time Transient Survey \citep[CRTS;][]{Drake09} light-curve, shown in Fig.~\ref{fig:lightcurve}, indicates very little fluctuation in the $V$-band magnitude and hence in the continuum flux. In addition, no significant variation in the UV continuum is seen in the COS data. We found a maximum of only $\sim$30 per cent variation in the UV flux between Epoch-$1^\prime$ and Epoch-3. Using the \MgII\ emission line width, \cite{Chand10} have estimated a black hole mass of $M_{\rm BH}\sim$$10^{9} M_{\odot}$ for this QSO which corresponds to a Schwarzschild radius of $R_{\rm Sch}\sim$$3\times10^{14}$~cm. Following \cite{Hall11}, we estimate the diameter of the disk within which 90 per cent of the 2700 \AA\ continuum is emitted, $D_{2700}\sim$50$R_{\rm Sch}\sim$$1.3\times10^{16}$~cm\footnote{We note that the $M_{\rm BH}$ and the rest-frame 3000~\AA\ luminosity of the present source are, respectively, $\sim$3 times lower and $\sim$5 times higher than the one studied by \cite{Hall11}. The facts ensure that the disk temperatures, and hence the size of the continuum emitting regions, are consistent with each other at $\sim$15 per cent level.}. From the rest-frame 1350~\AA\ flux (i.e., $\sim$5$\times10^{-15}$~erg~cm$^{-2}$~s$^{-1}$~$\rm \AA^{-1}$), measured from the STIS spectrum, we obtain the broad line region (BLR) size of $R_{\rm BLR}\sim$0.7--0.9~pc by adopting the $R-L$(1350\AA) relation given in \cite{Eser15}.

\section{Discussion}   
\label{sec:diss}  

We report on the variability in \NeVIII\ absorption originating from a mini-BAL outflow from the QSO PG~1206+459 (\zem\ = 1.163). The absorber is detected at \zabs\ = 1.02854 corresponding to an outflow velocity of $\sim$19,200 \kms. The mini-BAL system, with \NeVIII\ absorption kinematically spread over $\Delta v \sim$360~\kms\ and with a total column density of $\log~(N(\NeVIII)/{\rm cm^{-2}}) > 16$, was first reported in M13 (see their Table~2). Here we report a $\sim$4$\sigma$ variation in the observed EWs of the \NeVIII~$\lambda770$ and $\lambda780$ absorption lines over a timescale of 2.8 months in the QSO's rest-frame. Such variations cannot be attributed to systematic uncertainties since the neighboring unrelated, intervening absorption do not vary with such a high significance. This is the first reported case of an EUV mini-BAL variability. However, we did not observe any considerable variation in the \NeVIII\ absorption troughs on a timescale of $\sim$3.2 days in the QSO's rest-frame. Additionally, two other low-ionization species (\OIV\ and \NIV) with multi-epoch observations do not exhibit any appreciable variation at any time, e.g., show $<2\sigma$ variations. Note that the mini-BAL absorbers in the sample of \cite{Misawa14} show $<2\sigma$ variability over a similar timescale. Thus a $\sim$4$\sigma$ absorption variability in \NeVIII\ over 2.8 months in the QSO's rest-frame is somewhat extreme. However, we point out that the mini-BAL absorbers in their sample are detected via \SiIV, \CIV, and \NV\ absorption lines which have ionization potentials similar to \NIV\ and \OIV. The lack of significant variability in \NIV\ and \OIV\ lines in the present system, therefore, is consistent with the observations of \cite{Misawa14}.

As mentioned in Section~\ref{sec:intro}, the change in ionization state of the absorber and the change in $f_c$ due to absorber's motions are the two most plausible reasons for absorption variability. Below we discuss the two scenarios in detail.

\subsection{Variability due to change in ionization}      
\label{subsec:ionization}      

Variation in the optical depth of an absorption line (i.e. variability) naturally occurs due to the change in ionization state of the absorbing gas. The change in ionization state of photoionized gas can be induced via a change in the ionizing continuum. In the ``screening gas" scenario of \cite{Misawa07a}, the change in optical depth of ``screening gas" can also lead to absorption line variability. Nevertheless, variation in the continuum flux is thought to be the most probable reason for the variation in the optical depth of ``screening gas". Here we have found that the $V$-band magnitude and$/$or the UV continuum of the background QSO do not show any significant variation (Section~\ref{subsec:QSOprop}). The weak variation (up to 30\%) seen in the UV continuum will not change the \NeVIII\ column density appreciably. This is because $N(\NeVIII)$ reaches at its peak around ionization parameter of $\log U \sim$1 and shows a very little change over a wide range in $\log U$ around the peak, i.e., $\log U\sim$0.5--1.5 (see Fig.~12 of M13 for example). Thus a maximum of 30\% change in $\log U$ induced by the change in the UV continuum, assuming the shape of the ionizing spectrum is not changing, will not be discernible. Furthermore, the total column density of \NeVIII\ as given in M13 (i.e. $\log~(N/{\rm cm^{-2}}) > 16$, after correcting for the partial coverage) indicates that the profiles are too heavily saturated to be susceptible to the change in the ionization state. Finally, the lack of any considerable variation in the ``true" \NeVIII\ EWs, as measured from the partial coverage corrected spectra, and the lack of any change in the wings of the \NeVIII\ profile (Section~\ref{subsec:ABSprop}) indicates that it is highly unlikely that the observed line variation is due to the change in ionization state of the absorbing gas.

\subsection{Variability due to absorber's motion}      
\label{subsec:motion}  

Motions of the absorbing gas cloud, transverse to the line of sight, could result in a change in $f_c$ and hence in the ``apparent" optical depth of absorption. In Section~\ref{subsec:ABSprop} we have shown that the $f_c$ of the gas phase giving rise to the \NeVIII\ absorption shows an increase from $f_c = 0.59\pm0.05$ to $0.72\pm0.03$ over a timescale of 2.8 months in the QSO's rest-frame. Such an increase in $f_c$ value can occur primarily because of the two reasons: (i) transverse expansion of the absorbing gas cloud, and$/$or (ii) absorber's bulk motions across the line of sight.         

\subsubsection{Expansion across the line of sight} 

If the entire absorber is already within our cylinder of sight then any transverse motion will lead to a decrease in $f_c$ in contrast to what we observe here. Nevertheless, a gas cloud, without a significant transverse motion, can expand in size and consequently increase the covering fraction. In this scenario, in Epoch-1, the transverse size of the absorbing gas cloud is estimated to be $r_{\perp}=\sqrt f_{c} D_{2700}/2 \sim$$5.0\times10^{15}$~cm. An increase in radius from $r_1$ to $r_2$, due to expansion, will increase the covering fraction from $f_1$ to $f_2$ via the relation: $r_1/r_2 = \sqrt{f_1/f_2}$. The fact that the \NeVIII\ covering fraction increases from $\sim$0.59 to $\sim$0.72 in the rest-frame 2.8 months allows us to estimate an expansion rate of $\sim$720~\kms. If this were adiabatic expansion at the speed of sound then it would imply a temperature of $\sim$$4\times 10^7$~K. But such a high temperature is incompatible with the ionization states of metals that we observe in the absorber.  

More than 14 absorption troughs from eight different low- (e.g. \OIV, \NIV, \NV) and high-(e.g. \NeVIII, \NaIX, \MgX) ions are detected from this system (see Fig.~7 of M13). Interestingly, in M13 we have shown that these ions exhibit ionization potential (IP) dependent covering fractions with higher ions covering more of the continuum emitting region. For example, the \NIV\ and \OIV\ have $f_c$ values of $\sim$0.4 and 0.2, respectively, as opposed to $\sim$0.6 we measured for the \NeVIII. Such observations clearly suggest multiphase structure of the absorber with different ions having different projected area across the line of sight. Different phases of a multiphase photoionized absorber are often found to have very different densities and temperatures causing pressure gradients in the absorbing gas \citep[e.g.][]{Muzahid12b}. Such pressure gradients could, eventually, lead to an expansion. 

It is evident from Fig.~12 of M13 that $N(\NeVIII)$ peaks at $\log U =$~1.0, whereas, $N(\OIV)$ and $N(\NIV)$ peaks at $\log U =$~$-1.0$. Note that the photoionization equilibrium temperatures for the high- and low-ionization phases are $T\sim$$10^{5}$ and $\sim$$10^{4}$~K, respectively. At these temperatures, the gas-pressures for both the phases are at least 2 orders of magnitude higher than the corresponding radiation pressures, for the whole range of densities. Therefore, the total pressure, dominated by the gas-pressure, of the low-ionization phase is a factor of $\sim$10 higher than that of the high-ionization phase. As a consequence, it is natural for the low-ionization gas phase to exhibit a more rapid expansion than the \NeVIII\ bearing gas. However, we do not find any appreciable variability in the \NIV\ and$/$or \OIV\ absorption. Hence, the expansion scenario, although highly plausible in case of a multiphase absorber, cannot explain our observations. In view of the above arguments, we disfavor the expansion interpretation for the change in $f_c$ in this system.

\begin{figure*} 
\centerline{\vbox{ 
\centerline{\hbox{ 
\includegraphics[width=0.90\textwidth,angle=00]{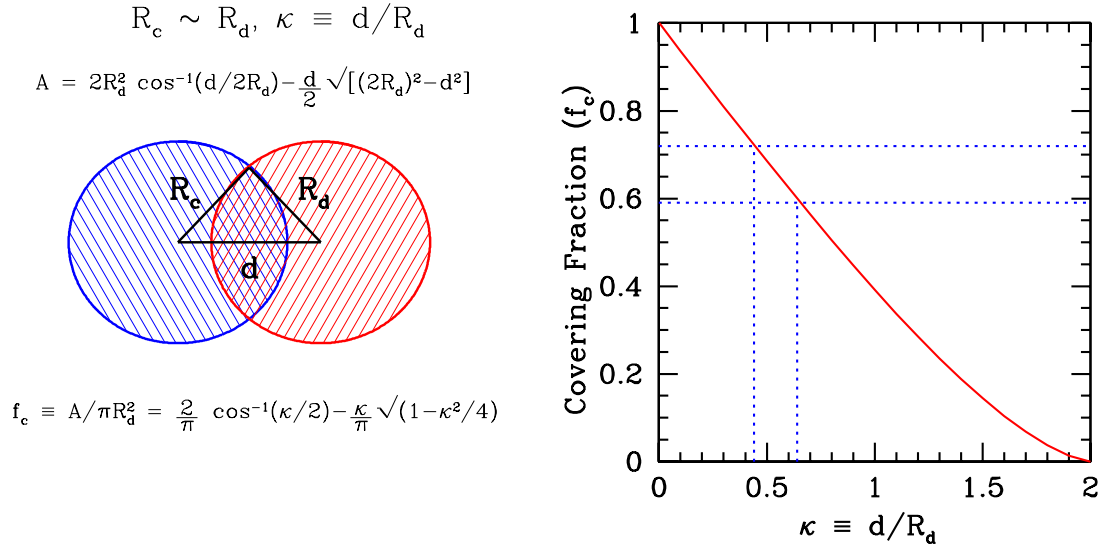}  
}}
}}
\vskip-0.35cm 
\caption{{\sl Left:} A simple model of a transiting cloud. The cloud with a radius 
$R_c$ is crossing the disk with a radius $R_d$. The cloud's center is separated by a  
distance of $d$ from the center of the disk. The covering fraction of the cloud is 
the ratio of the area common to both the circles ($A$) to the total area of the disk.  
{\sl Right:} The covering fraction as a function of $\kappa \equiv d/R_d$,   
assuming $R_c \sim R_d$.   
} 
\label{fig:geometry}    
\end{figure*}   

\subsubsection{Bulk motions across the line of sight} 

In a scenario in which an absorber is entering to our cylinder of sight, due to a rotational motion around the central SMBH, can explain the phenomenon of increase in $f_c$ with time. In Fig.~\ref{fig:geometry} we present a simple model in which a cloud with a radius of $R_c$ is transiting a continuum emitting disk with a radius of $R_d$. Since the absorber covers $\sim$60--70\% of the background source, we reasonably assume $R_c \sim R_d$. The common area between the cloud and the disk, under such an approximation, can be expressed as   
\begin{equation}  
A = 2 R_d^2~{\rm cos}^{-1}(d/2R_d) - \frac{1}{2} d~\sqrt{(2R_d)^2-d^2},     
\end{equation}    
leading to a covering fraction of    
\begin{equation}  
f_c = \frac{2}{\pi}~{\rm cos}^{-1}(\kappa/2) - \frac{\kappa}{\pi}~\sqrt{1-\kappa^2/4}~.      
\end{equation}    
Here, $d$ is the center-to-center projected separation between the disk and the absorber, and $\kappa$ is a dimensionless parameter defined as $\kappa \equiv d/R_d$. Note that $d$ can vary from 0 to $2R_d$ and $\kappa$ can have values between 0 and 2. The right panel of Fig.~\ref{fig:geometry} shows how $f_c$ changes with $\kappa$. The observed increase in $f_c$ value from 0.59 to 0.72, as shown by the horizontal dotted lines, corresponds to $\Delta \kappa \sim$0.2 leading to $\Delta d$~$\sim$$0.2R_d = 0.2D_{\rm 2700}/2 \sim$$1.3\times10^{15}$~cm. The transverse speed of the absorber is therefore,  $v_{\perp} = \Delta d/\Delta t \sim$1800~\kms. Note that the transverse speed is $\sim$10 times smaller than the radial outflow velocity we measured from the spectra (i.e. $\sim$19,200~\kms). If we assume that the $v_{\perp}$ can be matched to a Keplerian velocity \citep[see e.g.,][]{Mcgraw15}, then the derived distance between the absorber and the central engine is $r_d \sim$$4.0\times10^{18}$~cm $\sim$1.3~pc. This suggest that the outflowing gas is located just outside the BLR region. With the observed radial velocity the cloud will move only $\Delta r_d \sim$$4.5\times10^{-3}$~pc in the radial direction over the course of the 2.8 months. Clearly, the $\Delta r_d$ is too small, as compared to the $r_d$, to cause any significant change in the photon flux due to the change in the distance between the QSO and the absorber.            

A detailed photoionization model for this absorber was presented in M13. Here we recall the important model parameters: i.e., $\log U\sim$1, metallicity of $\rm [X/H] \sim$1.0, and total hydrogen column density of $\log~(N_{\rm H}/{\rm cm^{-2}}) = 20.7$. Inserting $r_d = 1.3$~pc in Eq.~4 of M13\footnote{$\log~\left(\frac{n_{\rm H}}{10^{5}\rm~cm^{-3}}\right) = \log~L_{912 \rm~\AA}^{30} - \log \left(\frac{r_d}{100 \rm~pc}\right)^2 - \log U - 1.25$, where $\log L_{912 \rm~\AA}^{30} = 32.1$, is the monochromatic luminosity of the QSO at the Lyman continuum in units of $10^{30}$~erg~s$^{-1}$~Hz$^{-1}$.}, we calculate the density of the \NeVIII\ bearing gas to be $n_{\rm H} \sim$$5.3\times10^{6}$~cm$^{-3}$. Given the density, the line of sight thickness of the cloud turns out to be $\delta r = N_{\rm H}/n_{\rm H}$ \linebreak $\sim$$9.5\times10^{13}$~cm. On the other hand, since the absorber covers at least $\sim$60 per cent of the continuum emitting region, the minimum transverse size of the absorber is $r_{\perp} = \sqrt f_c D_{2700}/2 \sim$$5.0\times10^{15}$~cm. Evidently, $r_{\perp} > 50~\delta r$, which is suggestive of a ``sheet-like" geometry. 

In a standard thin-shell model, the kinetic luminosity of an outflow can be expressed as $\dot{E}_{k} = 4\pi\mu m_{p} C_{\Omega} C_{f} N_{\rm H} r_d v^3$, where, $m_p$ is the proton mass, $\mu=1.4$ accounts for the mass of helium, and $C_\Omega$ and $C_f$ are, respectively, the global and local covering fractions \citep[see Appendix~C of][]{Muzahid15b}. In M13 we have shown that 40 per cent of the intermediate redshift QSOs exhibit highly ionized outflows detected via the \NeVIII\ absorption. Therefore, $C_\Omega$ can be taken as 0.4 for the general population. However, we note that the outflow velocity ($\sim$19,200~\kms) we measure for this system is very rare. Only one out of 20 QSOs shows such a high outflow velocity. We, thus, conservatively use $C_\Omega = 1/20$ for our calculations. The $C_f$ is related to the clumpiness of the absorbing material. For simplicity, $C_f$ is taken to be unity for the highly ionized diffuse gas. Using the distance $r_d\sim$1.3~pc, derived from our simple model, we estimate a kinetic luminosity of $\dot{E}_{k}\sim$$2.1\times10^{43}$~erg~s$^{-1}$. The $\dot{E}_{k}$ value increases to $\sim$$1.7\times10^{44}$~erg~s$^{-1}$ if we assume $C_\Omega = 0.4$. Note that the $\dot{E}_{k}$ of $\sim$$10^{43-44}$~erg~s$^{-1}$ is among the highest kinetic luminosities measured till date \citep[see Table~10 of][]{Dunn10}. Such large kinetic luminosities suggest that the outflows detected via the \NeVIII\ absorption are potentially major contributors to AGN feedback. 

It is now important to address whether we could explain the lack of variability in the low-ionization lines, i.e. \OIV\ and \NIV, with our simple model. Naively one would expect that the low-ionization gas phase would trace a high density, compact region. In M13, using \OIV~$\lambda608$ and $\lambda787$ transitions, we determined the \OIV\ covering fraction to be $\sim$0.2. Therefore, the transverse size of the low-ionization gas phase, giving rise to \OIV\ (and \NIV) absorption, is only a factor of $\sim$2 smaller than the \NeVIII\ bearing gas. The fact that all the high- and low-ionization lines coincide in velocity, possibly, indicates that the gas phases are co-spatial with the low-ionization lines stemming from the core of the absorber with a lower $f_c$ value. While such a scenario is reasonable, the difference in transverse sizes is not quite significant.

We recall that according to Eq.~\ref{eqn:sig}, the change in EW, and hence in $S_{ij}$, depends on the EW itself. From Fig.~\ref{fig:vary} it is evident that the both \OIV\ (EW~$=0.37\pm0.05$~\AA) and \NIV\  (EW~$=0.74\pm0.03$~\AA) lines are much weaker than the \NeVIII\ (EW~$=1.18\pm0.04$~\AA). Thus, one possibility here is that variations in those weak lines, \OIV\ in particular, are below our detection limit. In order to check that, we estimated the $S_{ij}$ value for \OIV\ in the partial coverage corrected spectra. However, we did not find any significant variation. Higher S$/$N multi-epoch observations are essential for reaching a firm conclusion on the degree of variability of these ions. As noted earlier, the lack of appreciable variation in the low-ionization species over a timescale of 2.8 months, however, is consistent with the recent observations by \cite{Misawa14}.

\section{Summary} 
\label{sec:summ}   

We have studied the variability of the \NeVIII~$\lambda\lambda$770, 780 doublet originating from a mini-BAL outflow with an ejection velocity of $\sim$19,200~\kms\ from the UV bright QSO PG~1206+459 (\zem~$=$~1.163). This is the first case of mini-BAL variability detected in EUV transitions. The \NeVIII\ doublet shows variability with a $\sim$4$\sigma$ significance over a timescale of 2.8 months in the QSO's rest-frame. However, no significant variation is observed over a shorter timescale of 3.2 days. Additionally, we did not observe any appreciable variation in lower ionization \OIV\ and \NIV\ lines. 

We have explored three different possibilities for the observed variability: (a) change in the ionization conditions, (b) expansion of the absorber, and (c) bulk motion of the absorbing gas across the line of sight. From the lack of variability in the partial coverage corrected spectra and estimated large column density of \NeVIII\ ($N>10^{16}$~\sqcm) we rule out the possibility of a change in ionization conditions. The expansion scenario leads to an unreasonably high gas temperature that is inconsistent with the ionization state of the detected metal lines. Besides, we note that the low-ionization lines should exhibit a more rapid expansion than the \NeVIII\ in contrast to what we observed here. We favored a scenario in which the absorbing gas is entering into the cylinder of sight and hence showing an increase in covering fraction. Using a simple model of a transiting cloud, with a size on the order of the continuum emitting region, we derived the transverse speed, distance from the central engine, density, and kinetic luminosity of the outflow. In order to explain the lack of variability in the low-ionization lines, we preferred a multiphase structure of the absorbing gas in which the low-ionization species stem from the core of the absorber with lower $f_c$ values and presumably already in the cylinder of sight. Nonetheless, the transverse size of the low-ionization gas phase is only a factor of $\sim$2 smaller than that of the high-ionization phase, posing an apparent tension to our favored scenario. High S$/$N, multi-epoch observations are needed for a firm conclusion on the variability of the weak, low-ionization lines.

Eight different ions with a wide range in IPs (i.e. 40--400 eV) are detected in this absorber (see Fig.~7 \& 8 of M13). A study of both short-term and long-term variability with multi-epoch $\it HST$ observations, in future, for more than 10 absorption troughs would provide further insights on the dynamics and the structure of this intriguing mini-BAL outflow.

\vskip0.4cm 
\noindent  
{\it Acknowledgements:} We would like to thank the anonymous referee for constructive suggestions which improved the manuscript. SM thankfully acknowledges IUCAA (India) for providing hospitality where a part of the work was done. SM also thanks Chris Culliton for useful discussion and carefully reading an earlier version of the manuscript. JC and ME acknowledge support from grant AST--1312686 from the National Science Foundation.

\appendix  
\section{Summary of equivalent widths and $S_{\rm ij}$ measurements}       
\label{app:EWlist}  
\begin{table*}
\begin{center}  
\caption{Summary of EWs and $S_{\rm ij}$ measurements.}         
\begin{tabular}{ccccrccr} 
\hline 
$\lambda_1$ (\AA) & $\lambda_2$ (\AA) &  $\rm EW_1$ (\AA) &  $\rm EW_2$ (\AA) &  $S_{12}$   &$\rm EW_{1^\prime}$ (\AA) &  $\rm EW_3$ (\AA) &  $S_{1^\prime3}$  \\ 
(1)        &  (2)        &  (3)            &  (4)            &  (5)      &  (6)            &  (7)            &  (8)     \\ 
\hline       
1460.7441  &   1462.1398 &   0.23$\pm$0.02 &   0.21$\pm$0.02 &   $-$0.6  &  0.20$\pm$0.02 &   0.16$\pm$0.02  &  $-$1.3  \\ 
1463.0298  &   1464.7288 &   0.44$\pm$0.02 &   0.43$\pm$0.02 &   $-$0.5  &  0.45$\pm$0.02 &   0.42$\pm$0.02  &  $-$0.9  \\ 
1464.6884  &   1467.8639 &   1.27$\pm$0.03 &   1.26$\pm$0.03 &   $-$0.1  &  1.24$\pm$0.02 &   1.24$\pm$0.03  &     0.0  \\ 
1468.5472  &   1470.6407 &   0.85$\pm$0.03 &   0.79$\pm$0.03 &   $-$1.7  &  0.82$\pm$0.02 &   0.77$\pm$0.02  &  $-$1.7  \\ 
1471.0566  &   1472.1326 &   0.62$\pm$0.01 &   0.59$\pm$0.02 &   $-$1.4  &  0.64$\pm$0.01 &   0.62$\pm$0.01  &  $-$1.2  \\ 
1474.4214  &   1475.6957 &   0.91$\pm$0.01 &   0.89$\pm$0.02 &   $-$1.1  &  0.91$\pm$0.01 &   0.91$\pm$0.01  &  $-$0.5  \\ 
1475.6957  &   1477.9813 &   1.17$\pm$0.02 &   1.17$\pm$0.03 &      0.0  &  1.18$\pm$0.02 &   1.19$\pm$0.02  &     0.4  \\
1479.3688  &   1480.5581 &   0.13$\pm$0.02 &   0.19$\pm$0.02 &      1.8  &  0.13$\pm$0.02 &   0.15$\pm$0.02  &     0.7  \\
1481.4803  &   1483.7620 &   0.43$\pm$0.03 &   0.44$\pm$0.03 &      0.1  &  0.45$\pm$0.02 &   0.44$\pm$0.03  &  $-$0.3  \\
1484.2960  &   1486.0920 &   0.81$\pm$0.02 &   0.81$\pm$0.02 &   $-$0.0  &  0.82$\pm$0.02 &   0.85$\pm$0.02  &     1.2  \\
1501.3752  &   1502.8315 &   0.29$\pm$0.03 &   0.25$\pm$0.03 &   $-$1.1  &  0.29$\pm$0.02 &   0.26$\pm$0.02  &  $-$0.8  \\
1503.6245  &   1504.5574 &   0.22$\pm$0.02 &   0.23$\pm$0.02 &      0.4  &  0.24$\pm$0.01 &   0.20$\pm$0.02  &  $-$1.7  \\
1508.1309  &   1511.7717 &   0.37$\pm$0.04 &   0.39$\pm$0.05 &      0.3  &  0.40$\pm$0.03 &   0.38$\pm$0.04  &  $-$0.3  \\
1513.8568  &   1515.2911 &   0.26$\pm$0.03 &   0.20$\pm$0.03 &   $-$1.5  &  0.23$\pm$0.02 &   0.23$\pm$0.02  &     0.2  \\
1515.5824  &   1517.8801 &   1.47$\pm$0.02 &   1.44$\pm$0.03 &   $-$1.0  &  1.46$\pm$0.02 &   1.49$\pm$0.02  &     1.1  \\
1517.8801  &   1519.3525 &   0.99$\pm$0.02 &   0.99$\pm$0.02 &   $-$0.2  &  1.00$\pm$0.01 &   1.01$\pm$0.02  &     0.3  \\
1522.9287  &   1524.4659 &   0.42$\pm$0.03 &   0.44$\pm$0.03 &      0.4  &  0.42$\pm$0.02 &   0.38$\pm$0.03  &  $-$1.2  \\
1525.8251  &   1527.4271 &   0.64$\pm$0.03 &   0.62$\pm$0.03 &   $-$0.6  &  0.62$\pm$0.02 &   0.63$\pm$0.03  &     0.4  \\
1530.4724  &   1531.2815 &   0.21$\pm$0.02 &   0.15$\pm$0.03 &   $-$1.8  &  0.18$\pm$0.02 &   0.18$\pm$0.02  &  $-$0.3  \\
1532.8252  &   1534.3689 &   0.36$\pm$0.03 &   0.33$\pm$0.03 &   $-$0.7  &  0.36$\pm$0.02 &   0.38$\pm$0.03  &     0.4  \\
1536.3785  &   1538.2524 &   0.70$\pm$0.03 &   0.71$\pm$0.03 &      0.3  &  0.73$\pm$0.02 &   0.75$\pm$0.03  &     0.5  \\
1547.2039  &   1548.5242 &   0.38$\pm$0.03 &   0.43$\pm$0.03 &      1.1  &  0.38$\pm$0.02 &   0.40$\pm$0.03  &     0.4  \\
1548.8058  &   1551.0970 &   0.80$\pm$0.04 &   0.73$\pm$0.04 &   $-$1.1  &  0.74$\pm$0.03 &   0.69$\pm$0.04  &  $-$1.1  \\
1551.1261  &   1553.3009 &   0.78$\pm$0.04 &   0.72$\pm$0.05 &   $-$1.0  &  0.74$\pm$0.03 &   0.72$\pm$0.04  &  $-0.4^a$    \\
1561.0355  &   1564.2556 &   1.21$\pm$0.05 &   1.23$\pm$0.06 &      0.3  &  1.18$\pm$0.04 &   1.43$\pm$0.04  &     $4.3^b$  \\
1564.8542  &   1565.9384 &   0.31$\pm$0.03 &   0.28$\pm$0.04 &   $-$0.5  &  0.29$\pm$0.02 &   0.30$\pm$0.03  &     0.3  \\
1580.8251  &   1584.6600 &   1.48$\pm$0.06 &   1.27$\pm$0.09 &   $-$1.9  &  1.39$\pm$0.05 &   1.69$\pm$0.05  &     $4.0^c$  \\
1633.3333  &   1635.2345 &   0.65$\pm$0.04 &   0.66$\pm$0.05 &      0.1  &  0.69$\pm$0.03 &   0.63$\pm$0.04  &  $-$1.1  \\
1635.5581  &   1639.5630 &   1.18$\pm$0.07 &   1.07$\pm$0.08 &   $-$1.1  &  1.14$\pm$0.05 &   1.05$\pm$0.06  &  $-$1.1  \\
1644.7046  &   1647.4554 &   0.33$\pm$0.06 &   0.39$\pm$0.07 &      0.6  &  0.37$\pm$0.05 &   0.30$\pm$0.06  &  $-$1.0  \\
1652.1115  &   1655.0240 &   0.50$\pm$0.06 &   0.62$\pm$0.07 &      1.3  &  0.55$\pm$0.05 &   0.56$\pm$0.06  &     0.1  \\
1669.7450  &   1672.0913 &   0.90$\pm$0.05 &   0.83$\pm$0.05 &   $-$1.0  &  0.85$\pm$0.04 &   0.83$\pm$0.05  &  $-$0.3  \\
1693.5436  &   1694.5145 &   0.23$\pm$0.04 &   0.22$\pm$0.04 &   $-$0.1  &  0.24$\pm$0.03 &   0.20$\pm$0.03  &  $-$0.8  \\
1710.5782  &   1713.4303 &   1.03$\pm$0.07 &   1.07$\pm$0.07 &      0.5  &  1.05$\pm$0.05 &   1.14$\pm$0.06  &     1.2  \\
1718.1025  &   1720.0240 &   0.61$\pm$0.06 &   0.56$\pm$0.07 &   $-$0.5  &  0.60$\pm$0.05 &   0.61$\pm$0.06  &     0.2  \\
1726.7717  &   1728.6649 &   0.69$\pm$0.06 &   0.71$\pm$0.06 &      0.3  &  0.67$\pm$0.05 &   0.72$\pm$0.05  &     0.7  \\
1734.0371  &   1736.1002 &   0.90$\pm$0.06 &   0.98$\pm$0.06 &      0.9  &  0.92$\pm$0.05 &   1.00$\pm$0.05  &     1.2  \\
1739.0291  &   1743.6892 &   2.41$\pm$0.13 &   2.56$\pm$0.09 &      0.9  &  2.47$\pm$0.07 &   2.41$\pm$0.08  &  $-$0.6  \\
1750.1213  &   1751.5032 &   0.65$\pm$0.05 &   0.70$\pm$0.05 &      0.7  &  0.69$\pm$0.04 &   0.62$\pm$0.05  &  $-$1.1  \\  
\hline  
\end{tabular}  
\label{tab:ewsum} 
\end{center}  
\flushleft  
Notes-- List of 39 strong absorption lines that are detected in Epoch-1, Epoch-2, and Epoch-3 data with  
$\rm EW/\sigma_{EW}>5$. Columns 1 and 2 are the lower and upper bounds of wavelengths, respectively, within 
which EWs are measured. Columns 3 and 4 are the observed EWs in Epoch-1 and Epoch-2 data, respectively. 
Columns 6 and 7 are the observed EWs in Epoch-$1^\prime$ and Epoch-3 data, respectively. The 1$\sigma$ errors  
in EW measurements (i.e. $\sigma_{\rm EW}$) are calculated after incorporating the continuum placement 
uncertainties. Columns 5 and 8 are the $S_{ij}$ values as defined in Eq.~\ref{eqn:sig}. 
$^{a}$\NIV~$\lambda$765, $^{b}$\NeVIII~$\lambda$770, and $^{c}$\NeVIII~$\lambda$780 absorption lines 
from the mini-BAL outflow studied here.       
\end{table*}

\section{Full Spectrum of PG~1206+459}   
\label{app:fullspec}     
\begin{figure*}   
\centering 
\includegraphics[width=0.95\textwidth,angle=00]{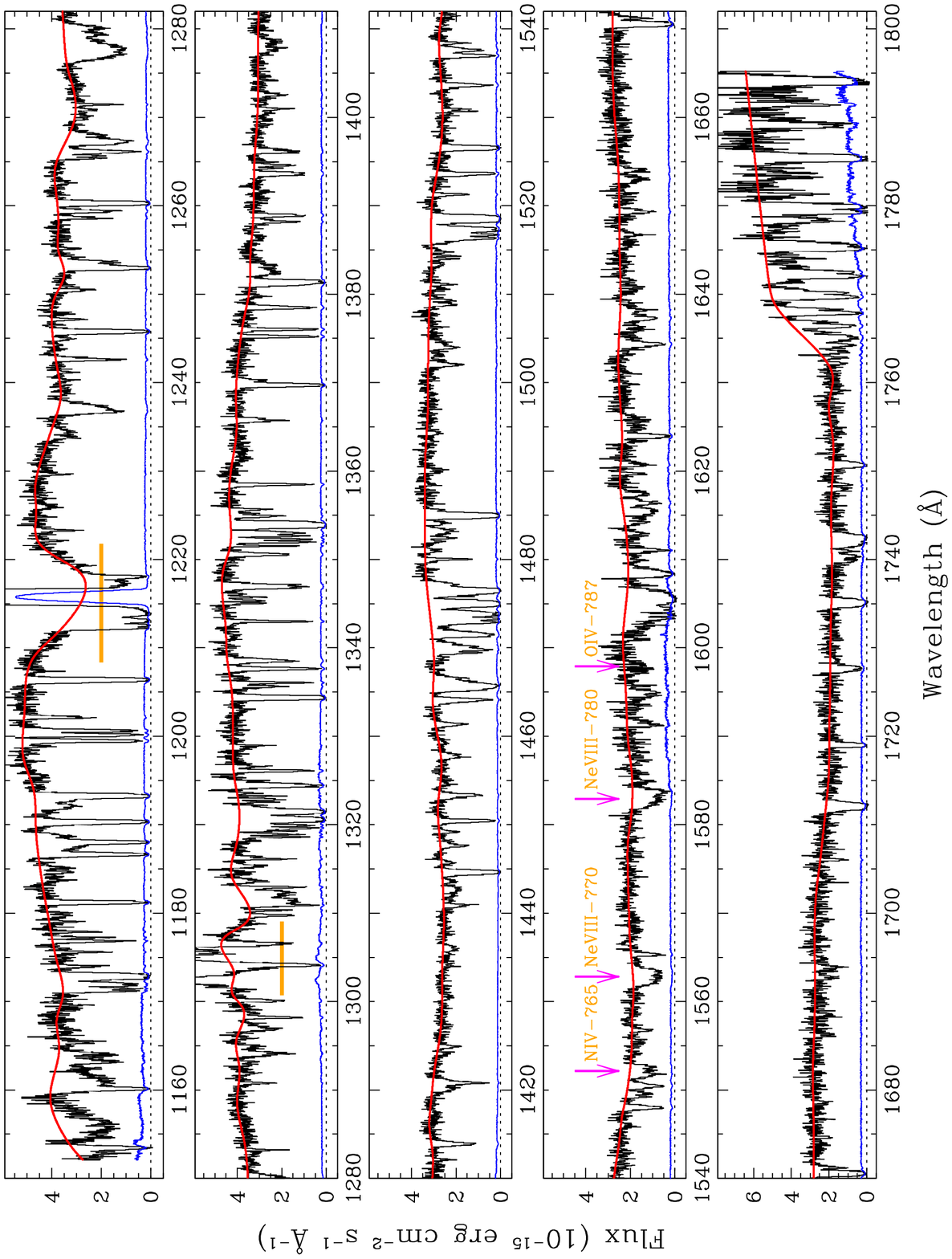} 
\vskip-0.3cm  
\caption{Epoch-$1^\prime$ spectrum of PG~1206+459 showing full wavelength coverage obtained using COS FUV (G130M+G160M) gratings. Error in each pixel is shown in blue. Continuum is shown by the smooth red curve. The absorption lines studied here are also marked. The wavelength ranges affected by the geo-coronal emission lines are indicated by the horizontal bars.}   
\label{fig:fullspec}  
\end{figure*}  

\def\aj{AJ}%
\def\actaa{Acta Astron.}%
\def\araa{ARA\&A}%
\def\apj{ApJ}%
\def\apjl{ApJ}%
\def\apjs{ApJS}%
\def\ao{Appl.~Opt.}%
\def\apss{Ap\&SS}%
\def\aap{A\&A}%
\def\aapr{A\&A~Rev.}%
\def\aaps{A\&AS}%
\def\azh{AZh}%
\def\baas{BAAS}%
\def\bac{Bull. astr. Inst. Czechosl.}%
\def\caa{Chinese Astron. Astrophys.}%
\def\cjaa{Chinese J. Astron. Astrophys.}%
\def\icarus{Icarus}%
\def\jcap{J. Cosmology Astropart. Phys.}%
\def\jrasc{JRASC}%
\def\mnras{MNRAS}%
\def\memras{MmRAS}%
\def\na{New A}%
\def\nar{New A Rev.}%
\def\pasa{PASA}%
\def\pra{Phys.~Rev.~A}%
\def\prb{Phys.~Rev.~B}%
\def\prc{Phys.~Rev.~C}%
\def\prd{Phys.~Rev.~D}%
\def\pre{Phys.~Rev.~E}%
\def\prl{Phys.~Rev.~Lett.}%
\def\pasp{PASP}%
\def\pasj{PASJ}%
\def\qjras{QJRAS}%
\def\rmxaa{Rev. Mexicana Astron. Astrofis.}%
\def\skytel{S\&T}%
\def\solphys{Sol.~Phys.}%
\def\sovast{Soviet~Ast.}%
\def\ssr{Space~Sci.~Rev.}%
\def\zap{ZAp}%
\def\nat{Nature}%
\def\iaucirc{IAU~Circ.}%
\def\aplett{Astrophys.~Lett.}%
\def\apspr{Astrophys.~Space~Phys.~Res.}%
\def\bain{Bull.~Astron.~Inst.~Netherlands}%
\def\fcp{Fund.~Cosmic~Phys.}%
\def\gca{Geochim.~Cosmochim.~Acta}%
\def\grl{Geophys.~Res.~Lett.}%
\def\jcp{J.~Chem.~Phys.}%
\def\jgr{J.~Geophys.~Res.}%
\def\jqsrt{J.~Quant.~Spec.~Radiat.~Transf.}%
\def\memsai{Mem.~Soc.~Astron.~Italiana}%
\def\nphysa{Nucl.~Phys.~A}%
\def\physrep{Phys.~Rep.}%
\def\physscr{Phys.~Scr}%
\def\planss{Planet.~Space~Sci.}%
\def\procspie{Proc.~SPIE}%
\let\astap=\aap
\let\apjlett=\apjl
\let\apjsupp=\apjs
\let\applopt=\ao
\bibliographystyle{mn}
\bibliography{mybib}

\end{document}